# Experiment Anti-Helium
## (Production of Nuclei and Anti-Nuclei.
## Limits for "exotic" particles of long lifetime )

Giorgio Giacomelli
University of Bologna and INFN Bologna
giacomelli@bo.infn.it

**Abstract**.
Data are recalled on the relative yields of $\pi^\pm$, $K^\pm$, $p^\pm$, $d^\pm$, $t^\pm$, $He^{3\pm}$, produced at 0° by 200-240 GeV/c protons on Beryllium and Aluminium targets. A search for the production of long-lived particles with charges 2/3, 1, 4/3, is described; For negative particle production the upper limits obtained at the 95 % Confidence Level were at the level of $10^{-11}$ with respect to the production of known particles($\pi^\pm$, $K^\pm$, $p^\pm$).

# 1. Introduction

In the 1970s CERN prepared in the West Area of the SPS a high intensity RF separated beam, with superconducting RF1-RF2 radiofrequencies, [1]. The beam was equipped with long gas threshold and differential Cherenkov counters (DISCs) to be used for several purposes [2].

A small collaboration of physicists from Annecy, Bologna and Saclay (10 physicists in total) organized a study of particle production at zero degrees ($\pi^\pm$, $K^\pm$, $p^\pm$ in proton-Berillium and proton-Aluminium collisions at 200-240 GeV/c [4,5,6,7] and compared the experimental results with the predictions of thermodynamic models [3]. They also made a precision search for anti-nuclei (specifically anti-$He^3$ e anti-$t^3$ for which several hundred events were recorded, several thousands for antideuterons). They also obtained new upper limits at the 95 % CL for the forward production of "exotic" particles with masses between 0.4-10 GeV [4-7].

Following previous measurements of anti-deuterons at Brookhaven [8] and CERN [9], it was interesting to study heavier anti-nuclei and place stronger limits on exotic particles.

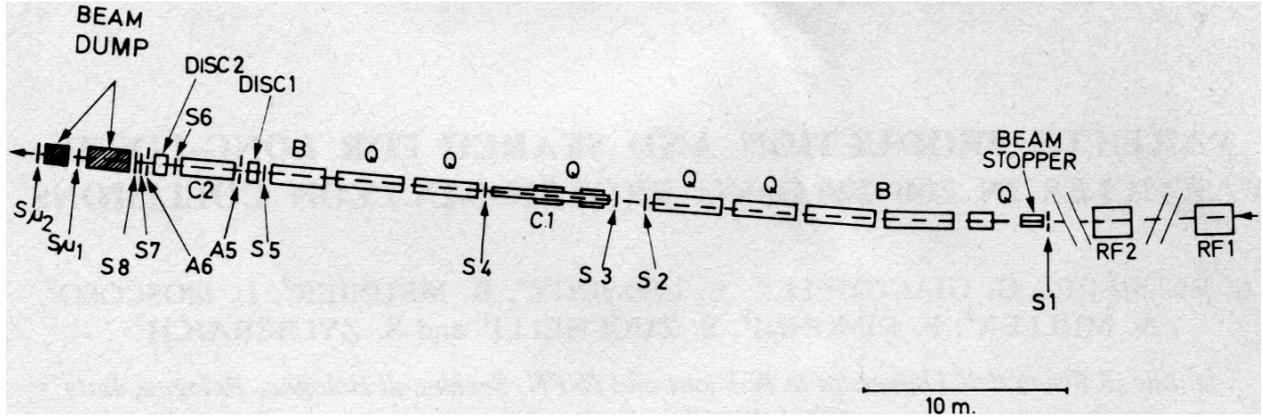

Fig. 1. The high intensity separated S1 beam (from right to left). S1-S8, A5-A6, Sm1-Sm2 are scintillation counters; C1-C2 are gas threshold Cherenkov; DISC1-DISC2 are gas Cherenkov differential counters.

## 2. The CERN S1 beam

The 200-240 GeV primary proton beam with an intensity of $\approx 10^{12}$ p per pulse hit a Be (Al) target 4-10 cm long. Secondary particles produced at $\approx 0°$ were analyzed in the S1 low momentum beam.

Fig. 1 shows the layout of the S1 beam in the SPS West Area. Incoming particles were defined by 8 scintillation counters of different sizes and thicknesses, S1-S8. Two large scintillation counters, $S_{\mu 1}$, $S_{\mu 2}$, were added downstream, behind a 3.6 m long iron beam dump.

Two gas threshold Cherenkov counters, (C1 and C2) vetoed high velocity particles, beam halo particles and particle interactions.

Two differential gas Cherenkov counters of the DISC type [2] were located 210 m from the primary targets (Be and Al). At the DISC position the counters S5 and S6 defined a beam size of 3 cm diameter, which corresponds to the DISC window aperture.

The main trigger was the coincidence of 8 scintillators vetoed by A5-A6 and C1 or C2.

The outputs of one or two DISCs were selected as needed.

Pulse Heights and times of flights for all scintillators and Cherenkov counters were recorded. The measured time accuracy of each scintillator was ±0.15 ns.

The superconducting separators removed high velocity particles, thus allowing effective acceptances of more than $2 \cdot 10^7$ particles per burst produced at the targets, while keeping the number of particles in the detectors at the $10^6$ level, which allowed proper measurements in the Cherenkovs and in the time of flight system and to achieve rejection ratios of better than $10^{-11}$. For 200 GeV proton- nucleon collisions the DISC Cherenkovs covered the pseudorapidity range 6<$\eta=\beta\gamma$<30, while the time of flight system covered the range 4<$\eta$<15. Thus for 4< $\eta$ <15 the velocity was measured by the DISC counters and by the time of flight system, and thus the identification power was maximal.

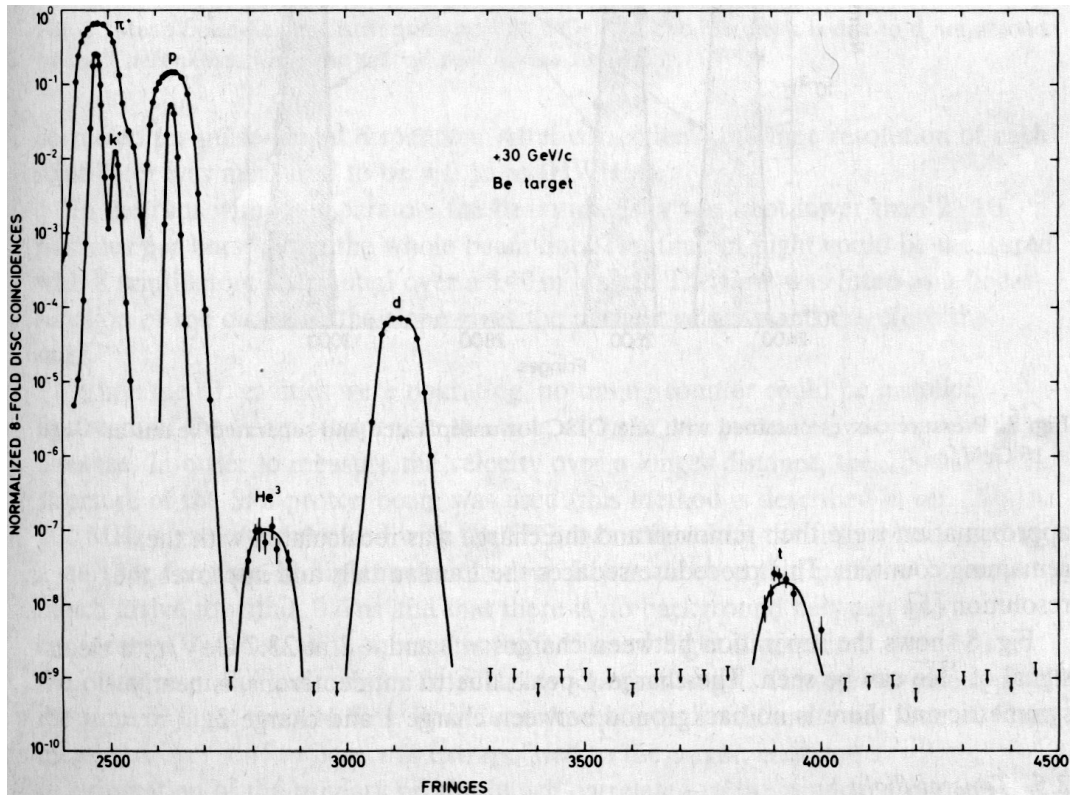

Fig. 2. The measurement of $\pi^+$, $K^+$, $p^+$ at 0° using the best resolution of one of the Cherenkov differential counters (DISCs with two different aperture selection) and of the time of flight. Note also the very good measurements of He$^3$, d, t.

## 3. Production of $\pi^\pm$, $K^\pm$, $p^\pm$

Fig. 3 shows the results of the measured particle ratios (measured with one of the DISC counters at high resolution) $K^+/\pi^+$, $p/\pi^+$, $K^-/\pi^-$, $p^-/\pi^-$ produced in Be and Al targets, compared to the predictions of thermodynamic models [3].
There are some disagreements between data and predictions; the global agreement was only correct as a general tendency.
The prediction of the dependence on the mass was an exponential dependence, see Fig. 7.

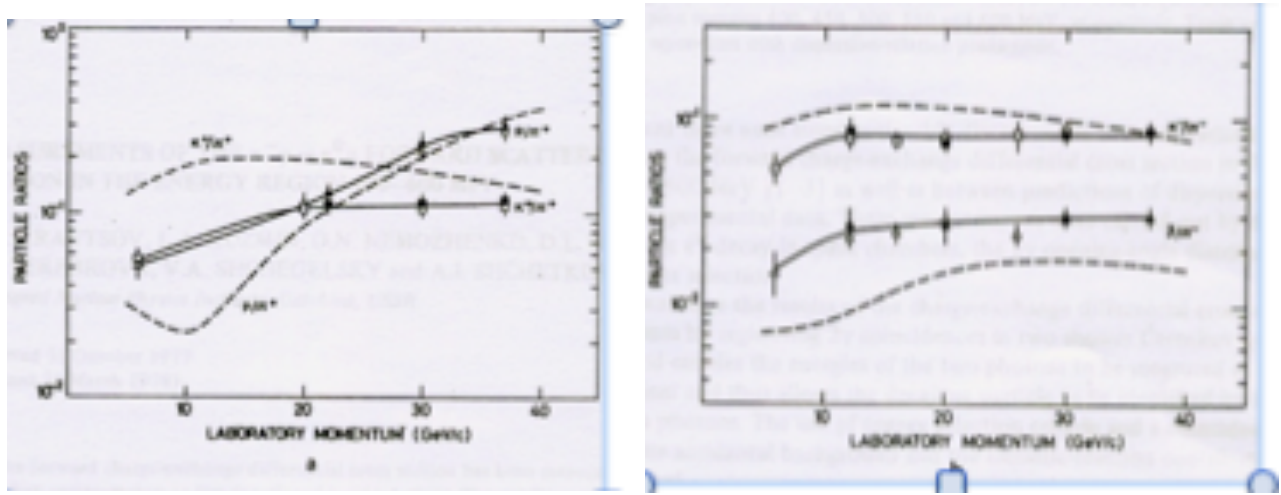

Fig. 3. Particle production ratios in Be (open points) and in Al (black points) compared to thermodynamic model predictions (dashed lines). (Left) Ratios $K^+/p^+$ and $p/p^+$; (Right) Ratios $K^-/p^-$ and $p^-/\pi^-$.

## 4. Production of nuclei and antinuclei

The ratios (nuclei/$p^+$) and (antinuclei/$p^-$) were measured at 200, 210, and 240 GeV/c protons in the Be target, while for the Al target only 200 GeV protons were used. The data were corrected for absorption and decay to the centers of the targets.

Fig. 4 shows the detection of anti-d [8, 9] and of anti-He3 at -23.7 GeV/c performed with the DISC counters in protons of 240 GeV/c –Be collisions. Notice the rather large number of anti-deuterons collected.

Fig. 5 shows the results of the detection of anti-d, anti-t and anti-He$^3$ [10-13] antinuclei using both the DISC Cherenkov counters and the pulse heights of the scintillation counters : this arrangement improves the selection and the rejection power of unwanted events.

Fig. 6 shows a compilation of the forward production of nuclei and antinuclei in 200-240 GeV/c proton collisions with Be and Al targets.

Fig 7 shows the invariant cross sections at x=0 plotted versus mass. The lines are fits to an exponential form of the type : $f=f_0 \exp(-bm)$. For both nuclei and anti-nuclei the dependence is of an exponential type with $f_0=(9.8\pm1.6)\times10^{-24}$ cm$^2$ GeV$^{-2}$ c$^3$ and a slope $b=(9.6\pm0.1)$ for antinuclei, and $f_0=(5.4\pm0.9)$, $b=(7.9\pm0.1)$ for nuclei. The comparison with the production of nuclei and anti-nuclei at other energies is difficult to make because of the scarcity of data and/or because they correspond to different kinematic conditions [10-21].

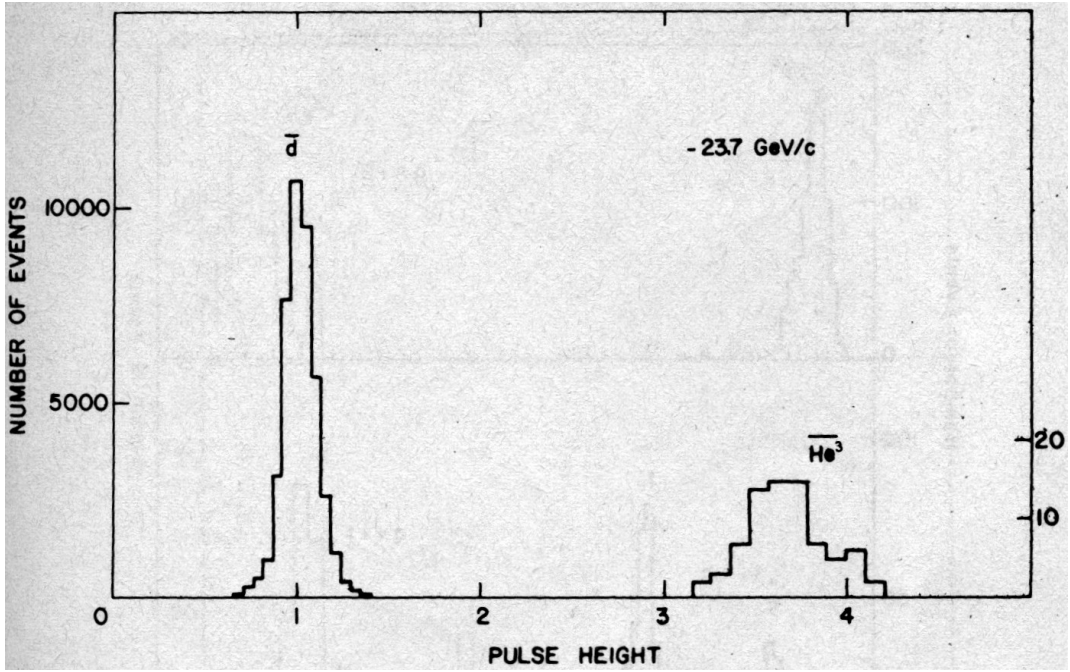

Fig. 4. The detection of anti-d and anti-He³ at -23.7 GeV/c performed with the DISC counters in (protons of 240 GeV/c –Be) collisions.

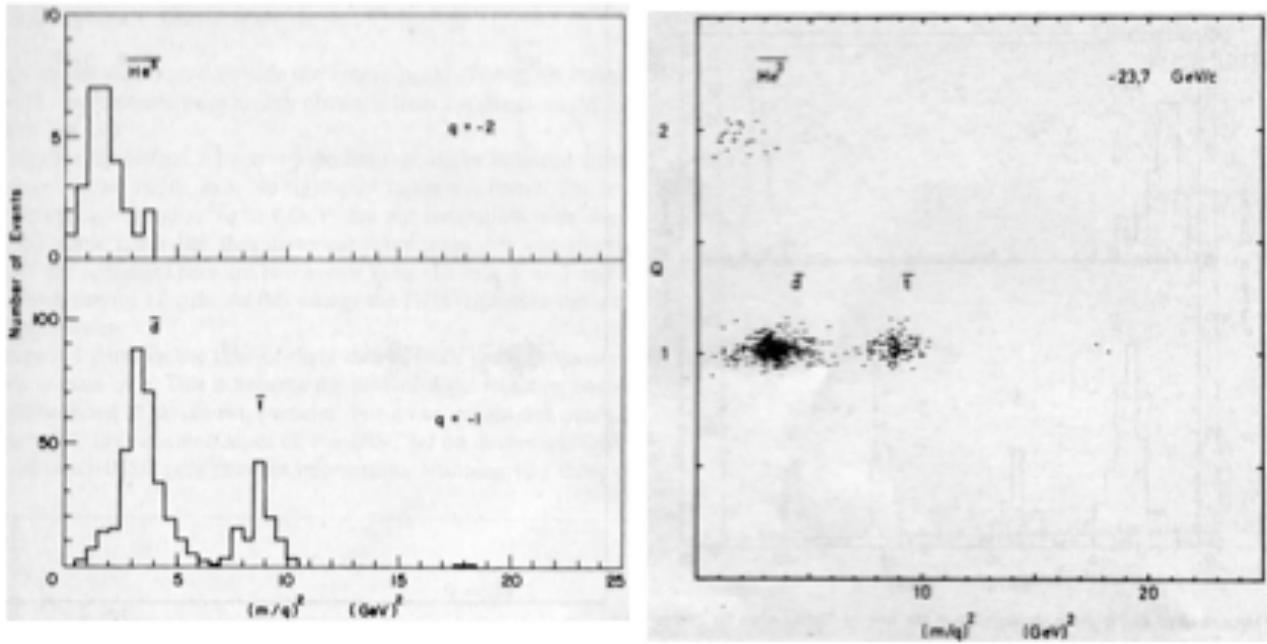

Fig. 5. Production of antinuclei using the DISC differential Cherenkov counters and the pulse heights in the scintillation counters.

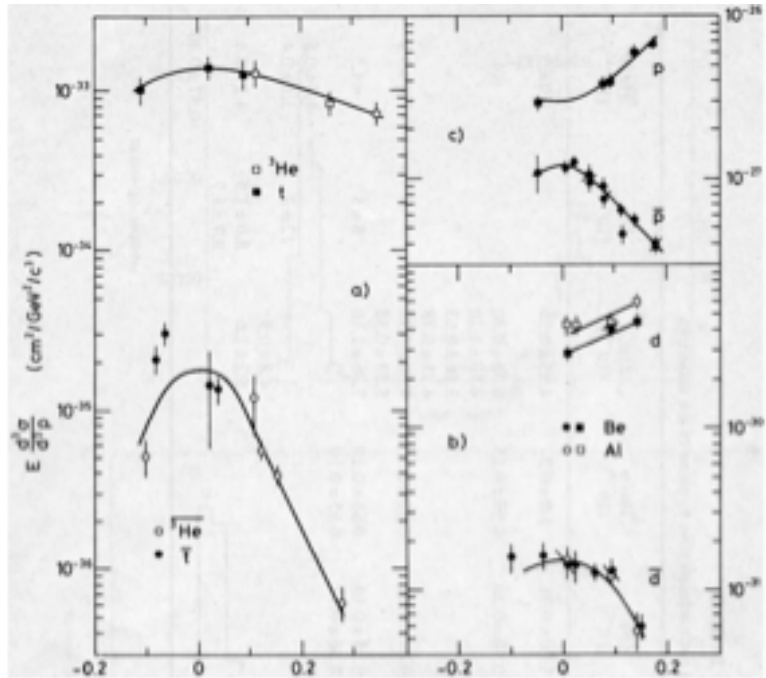

Fig. 6. Forward production of nuclei and antinuclei in protons of 200-240 GeV/c in collisions with Be and Al targets.

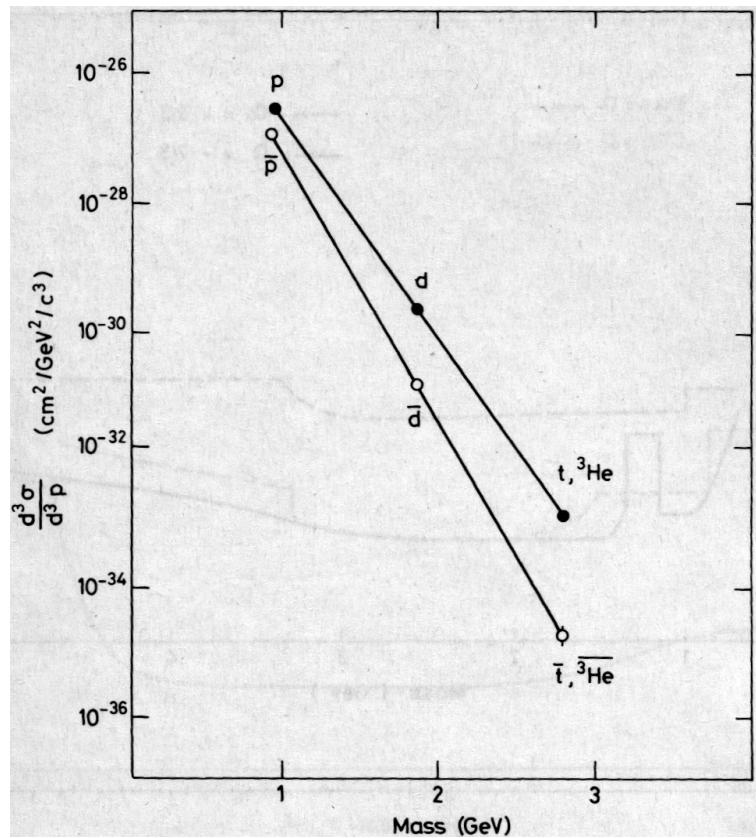

Fig. 7. Invariant cross section at x=0 for the production of nuclei and anti-nuclei plotted versus mass. The lines are fits to an exponential form, see text.

# 5. Upper limits for new long-lived particles

A careful search was made for new stable or quasi stable particles at each momentum. The combined measurement of charge and mass, as shown in Fig. 5 fo 23.7 GeV/c, was used to obtain new limits for unobserved new particle production. The mass squared, measured via time of flight, has a resolution of 1 GeV$^2$. This leads to good limits as long as there is no background.

For charges ±1 and ±2 the time of flight method is acceptable to establish limits for masses larger than 3 GeV. The DISC counters scanned a mass range in the vicinity of anti-t, anti-He3 and anti-d. The rejection power of a DISC could be increased by reducing its angular acceptance. The two methods were used at all momenta.

For particles with charges ±2/3 and ±4/3 there was no background; thus the limits were obtained from the measurements of charge and time of flight, see Figs. 8-9.

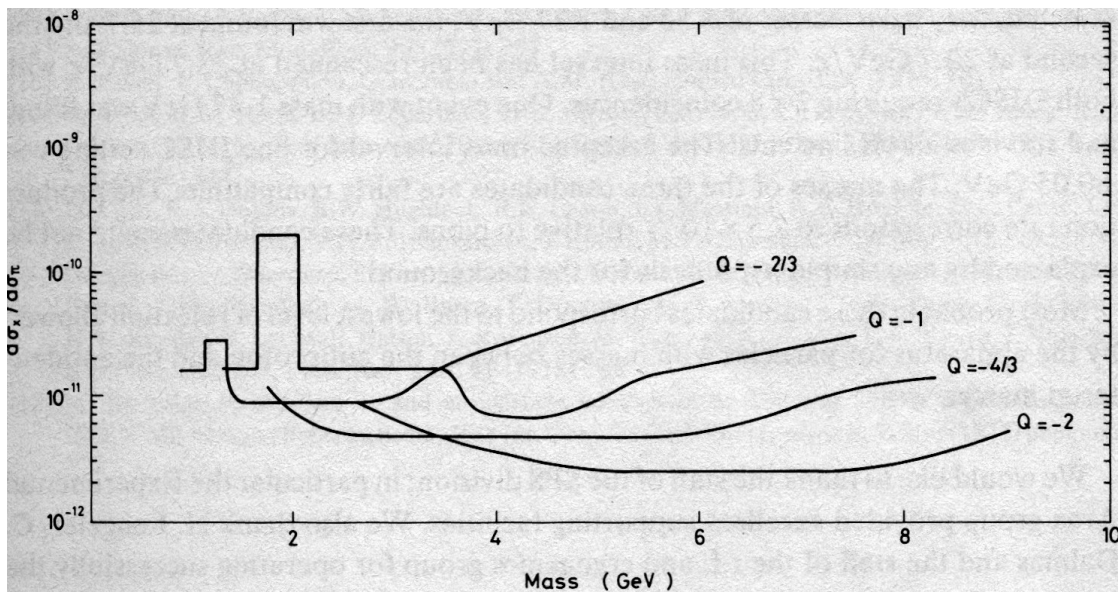

Fig. 8. Upper limits at 95 % CL for the production of long-lived heavy "exotic hadronic particles" with charges -2/3, -1, -4/3, -2 plotted vs particle mass.

The above limits for charged hadrons can be improved for long-lived charged heavy leptons. For part of the running time the rates $S_{\mu 1}$, $S_{\mu 2}$ were recorded. They allowed an extra good rejection also close to the known hadron masses.

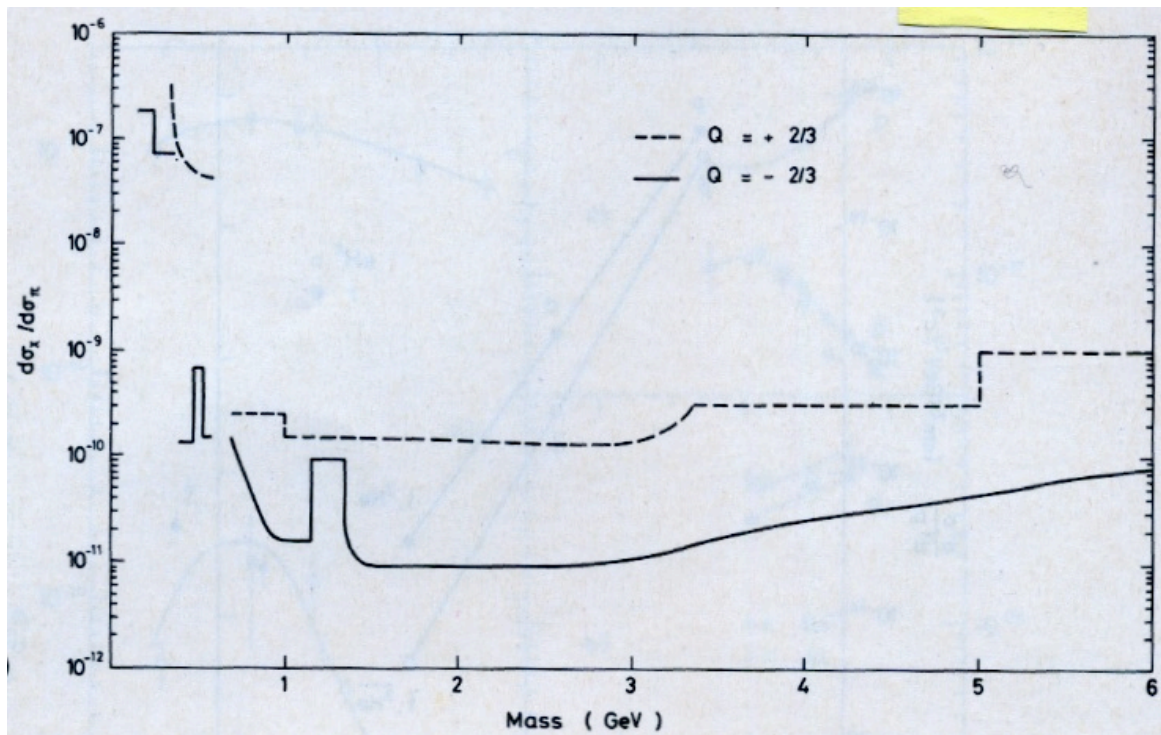

Fig. 9. Upper limits at 95 % CL for the production of long lived or quasi long lived leptonic new particles with charges +2/3 and -2/3. Notice that the upper limits for negative particles of charges -1, -2/3, are about one order of magnitude smaller ($10^{-11}$) than for positive particles ($10^{-10}$).

# 6. Conclusions

In a small, refined experiment, made by a small collaboration of 10 physicists of 3 different Institutions, using a high intensity beam separated with superconducting RF at the SPS of CERN were measured, in 1977-1980, a large number of anti-deuterons, and were found anti-nuclei of anti-$He^3$ and of anti-t .

Also the forward production cross sections at zero degrees for $\pi^\pm$, $K^\pm$, $p^\pm$ were measured, see Fig. 3.

The strong dependence of the forward production for nuclei and anti-nuclei was determined : it is of an exponential type for light nuclei and anti-nuclei till $He^3$, t, anti-$He^3$ , anti-t , see Fig. 7. Some of these antnuclei were observed at IHEP (Serpukhov) [10-13]. Anti-$He^4$ antinuclei were recently observed at BNL [22].

New upper limits were established for the forward production of different exotic long-lived states ( quarks, diquarks, …. ), from about 0.4 GeV to about 10 GeV mass. The best limits concern leptonic negative states, see Fig. 9, using in the trigger also the two scintillation counters $S_{\mu 1}$, $S_{\mu 2}$, placed at the end of the beam, immediately after the iron beam stopper.

## Acknowledgements

I would like to thank all the collaborators of the experiment, the CERN colleagues who prepared the SPS beam (D.E. Plane and H. Lengeler) [1], the colleagues who developed the Cherenkov differential counters [2] , and the technical staff, in particular  J. C. Bertrand, C. Lechauguette and A. Maurer. I thank dr. M. M. Deninno for the perfect organization of the Symposium.

## Bibliography

1. D. E. Plane, CERN/SPS/EA/78-1 (1976).
   H. Lengeler and D. E Plane, SPS/EPB/79-16.
   A. Citron, et al., Nucl. Instrum. Meth. 164 (1979) 31.
2. R. Meunier, et al., Nucl. Instr. 17 (1962) 1.
   M. Benot, et al., Nucl. Instr. 105 (1972) 431.
3. R. Hagedorn, Phys. Rev. Lett. 5 (1960)276.
   Grote-Hagedorn-Ranft, Particle spectra, CERN Report (1970).
4. A. Bussiere, G. Giacomelli, E. Lesquoy, R. Meunier, L. Moscoso, A. Muller, F. Rimondi, S. Zucchelli, S. Zylberajch (Particle Production and Search for Long-lived Particles in 200 to 240-GeV/c Proton-Nucleon Collisions) Nucl. Phys. B174 (1980) 1.
5. W. Bozzoli,  et al. (Production of d, t, $He^3$, anti-d,and anti He-3 by 200 GeV protons) Nucl. Phys. B144 (1978) 317.
6. W.Bozzoli, et al (Production of p+-,K+-, p and anti-p by 200 GeV Protons) Nucl. Phys. B140 (1978) 271.
7. W. Bozzoli, et al (Search for long-lived particles in 200-GeV/c
   proton-nucleon collisions) Nucl. Phys. B159 (1979) 363.
8. D. E. Dorfan, et al., Phys. Rev. Lett. 14 (1965) 1003.
9. T. Massam, et al., Nuovo Cimento 39 (1965) 10.
10. Y. M. Antipov, et al, Nucl. Phys. B31 (1971) 235.
11. N. K. Vishnevskii, et al., Sov. J. Nucl. Phys. 20 (1975) 371.
12. Yu. B. Bushnin, et al, Phys. Lett. B29 (1969) 500.
13. G. F. Binon et al, Phys. Lett. B31 (1970) 230.
14. M. Antinucci, et al., Lett. Nuovo Cim. 6 (1973) 121.

15. E. Albini, et al, Nuovo Cim. A32 (1976) 101.
16. W.F. Baker, et al, Phys. Lett. B51 (1974) 303.
17. P. Capiluppi, et al., Nucl. Phys. B79 (1974) 189.
18. G. Giacomelli, arXiv:0802.2241[hep-ph] (2008).
19. G. Giacomelli, arXiv:0712.0906[hep-ex] (2007).
20. A. Bertin, et al. Phys. Lett. 418 (1972) 201.
21. S. Carrol, et al., Phys. Lett. B80 (1979) 319.
22. Y. G. Ma, et al., arXiv:1301.4902 [nucl-ex] (2013).